# Phase diagram of pressure-induced high temperature superconductor La$_3$Ni$_2$O$_{7+\delta}$

Yuta Ueki[1,2]*, Hiroya Sakurai[1], Hibiki Nagata[1,2], Kazuki Yamane[1,2], Ryo Matsumoto[1], Kensei Terashima[1], Keisuke Hirose[3], Hiroto Ohta[3], Masaki Kato[3], Yoshihiko Takano[1,2]

[1]*National Institute for Materials Science (NIMS)*
*1-2-1 Sengen, Tsukuba, Ibaraki 305-0047, Japan*
[2]*University of Tsukuba 1-1-1 Tennodai, Tsukuba, Ibaraki 305-8577, Japan*
[3]*Doshisha University 1-3 Tataramiyakotani,*
*Kyo-Tanabe, Kyoto 610-0321, Japan*

We successfully synthesized samples of La$_3$Ni$_2$O$_{7+\delta}$ ($\delta$ = −0.50, −0.16, 0.00, +0.01, and +0.12) and measured the resistance under extremely high pressures using a diamond anvil cell to establish the electronic phase diagram. A Mott insulating state appears at $\delta$ = −0.50, where all Ni ions are divalent. With increasing oxygen content, superconductivity appears at $\delta$ = 0.00 and higher, above approximately 25 GPa, passing through Anderson localization at $\delta$ = −0.16. The superconducting transition temperature ($T_c$) decreases with increasing pressures for both $\delta$ = 0.00 and +0.12, with the pressure dependence of $T_c$ being much stronger in the latter than in the former.

## 1. Introduction

Electronic phase diagrams are highly informative for studying superconductors.[1—7] They highlight the thermodynamic conditions suitable for superconductivity and indicate potential fluctuations that induce or compete with superconductivity. These diagrams serve as benchmarks for theoretical models of superconducting mechanisms. Their importance is especially pronounced under extremely high pressures, where experimental data is limited due to technical challenges.

Recently, La$_3$Ni$_2$O$_7$ was found to exhibit superconducting transitions at pressures above 14 GPa.[8] This discovery is significant due to its high transition temperature of 80 K and its structural similarity to high $T_c$ cuprates. The crystal structure of La$_3$Ni$_2$O$_7$ consists of double NiO$_2$ layers and La—O rock salt blocks,[9] as shown in Fig. 1. The Ni sites form a square lattice similar to the Cu square lattice in CuO$_2$ planes. Superconductivity on NiO$_2$ planes has also been observed in thin films of infinite-layer nickelates,[10] where it occurs at temperatures of 15 K or lower when the Ni valence is between +1.15 and +1.25.[11, 12] In these films, the Ni ion, coordinated by four oxygen ions, has 8.75—8.85 $3d$ electrons, making a single-band model of the Ni $3d_{x2-y2}$ orbitals appropriate.[13] However, in $La_3Ni_2O_7$, the number of $3d$ electrons in Ni ions is different, and a single-band model does not apply because the Fermi surface involves two types of electronic bands, characteristic of the $3d_{x2-y2}$ and $3d_{z2}$ orbitals.[14] This difference, along with the higher $T_c$, suggests a different superconducting mechanism for $La_3Ni_2O_7$ compared to thin films.

In $La_3Ni_2O_7$, the Ni ion has 7.5 $3d$ electrons, which is commensurate with its bilayered crystal structure. For instance, two Ni ions along the $c$-axis, shown in the circle in Fig. 1, share three $3d$ electrons between the $3d_{x2-y2}$ and $3d_{z2}$ orbitals. Likely due to this commensurability and the almost consistent $T_c$ values reported,[8, 15—20] the effect of oxygen nonstoichiometry on superconductivity has received less attention, similar to the case of $Sr_2RuO_4$. However, superconductivity is clearly dependent on band filling, which can be adjusted by varying the oxygen content. Our recent findings show that the superconductivity of $La_4Ni_3O_{10}$ is very sensitive to its oxygen content.[21] Therefore, we aim to establish a comprehensive phase diagram of $La_3Ni_2O_7$ that includes temperature ($T$), pressure ($P$), and oxygen content.

## 2. Experimental

Polycrystalline samples of $La_3Ni_2O_{7+\delta}$ were prepared by solid-state reaction from $La_2O_3$ and NiO. The oxygen content was initially estimated to be $\delta = +0.01$ for samples slowly cooled in flowing oxygen gas. Detailed procedures for synthesis and oxygen-content determination are described elsewhere.[15] The oxygen content was then increased to $\delta = +0.12$ using hot isostatic pressing under 1500 kgf/cm² with 20% $O_2$/Ar at 600°C for 2 hours. The excess oxygen atoms are likely located in the rock salt blocks, similar to $La_2NiO_{4.17}$.[22, 23] Conversely, the oxygen content was decreased by annealing samples in air at 500°C for 2 days, or in flowing 10% $H_2$/Ar for 12 hours at 250°C or 300°C, resulting in $\delta$ values of 0.00, −0.16, and −0.50, respectively. The oxygen atoms are likely removed from the site between two Ni atoms, as shown previously.[24]

Sample characterization was performed using powder X-ray diffraction (XRD) with Cu $K_\alpha$ radiation on a commercial diffractometer, MiniFlex600 (Rigaku), equipped with a one-dimensional high-speed detector, D/teX Ultra. Electrical resistance, $R$, for each sample was measured under various pressures using a diamond anvil cell (DAC) with a commercial PPMS (Quantum Design) system, from 300 K to 2 K. The sample was placed between two diamonds with cubic BN as the pressure medium and stainless steel as the gasket. Electrical terminals made of boron-doped diamond were printed on the surface of one diamond. Detailed DAC configuration is provided elsewhere.[15, 25]

## 3. Results and Discussion

XRD patterns for all samples are shown in Fig. 2. All peaks for $\delta = +0.01$, 0.00, and $-0.16$ were indexed assuming orthorhombic *Amam* symmetry.[9] The lattice parameters were estimated as follows: $a = 5.3903$Å, $b = 5.4464$Å, and $c = 20.507$Å for $\delta = +0.01$; $a = 5.3920$Å, $b = 5.4510$Å, and $c = 20.533$Å for $\delta = 0.00$; and $a = 5.4083$Å, $b = 5.4487$Å, and $c = 20.480$Å for $\delta = -0.16$. For $\delta = -0.50$, tetragonal *I4/mmm* symmetry was adopted, consistent with previous reports for $\delta = -0.65$.[24] The lattice parameters were $a = 3.874$Å and $c = 20.075$Å for $\delta = -0.50$. In contrast, for $\delta = +0.12$, both orthorhombic and tetragonal phases were observed, as shown in Figs. 2*b* and 2*c*. The orthorhombic 020 and 200 peaks of $La_3Ni_2O_{7.01}$ partially merged into a single peak at $2\theta = 32.9°$, while the orthorhombic 0,0,10 peak partially remained at the same position and partially shifted to a higher angle of 44.9°. The lattice parameters for these phases were estimated to be $a = 5.384$Å, $b = 5.453$Å, and $c = 20.48$Å for the orthorhombic phase, and $a = 3.857$Å and $c = 20.14$Å for the tetragonal phase. The lattice parameters of the orthorhombic phase agree well with those of $La_3Ni_2O_{7.01}$. Similar phase separation has been observed in $La_2CuO_{4+\delta}$.[26, 27]

The weight ratio of the two phases in the $La_3Ni_2O_{7.12}$ sample was estimated using the reference intensity ratio (RIR) method to be approximately 35% orthorhombic phase and 65% tetragonal phase. Assuming the oxygen content of the orthorhombic phase is the same as $La_3Ni_2O_{7.01}$, the oxygen content of the tetragonal phase can be calculated as $\delta = +0.17$. This estimation is reasonable, as the $\delta$ value aligns with that estimated for a commensurate superstructure of $La_2NiO_{4.17}$.[22] From a structural perspective, the chemical formulas of these compounds can be described as $(LaO_{1+\delta})(LaNiO_3)_2$ for $La_3Ni_2O_{7+\delta}$ and $(LaO_{1.17})LaNiO_3$ for $La_2NiO_{4.17}$, as mentioned earlier.

The *R—T* curves at various pressures for $\delta = 0.00$ and $\delta = +0.12$ are shown in Fig. 3. The resistance of each sample decreases with increasing pressure. The temperature dependence is partly negative, likely due to grain boundary scattering, especially above around 160 K, as seen in $La_4Ni_3O_{10}$. This is because $La_3Ni_2O_{7+\delta}$ ($\delta \geq 0$) is known to exhibit metallic temperature dependence at least above 160 K at ambient pressure.[9, 28—35] Below approximately 160 K, semiconducting behavior attributed to a kind of density wave formation has been reported at ambient pressure in some studies,[33—35] although the other studies have reported metallic behavior even below 160 K.

For δ = 0.00 and δ = +0.12, the resistance at pressures above 25.6 GPa and 26.1 GPa, respectively, shows a kink at around 80 K or lower, indicating the superconducting transition. The transition temperature, estimated as the crossing point of lines just above and below the kink, is summarized in Fig. 4. The pressure dependence of $T_c$ is clearly different for each sample; the decrease in $T_c$ for δ = 0.00 is much smaller than that for δ = +0.12. Interestingly, despite the robustness of $T_c$ against pressure, the decrease in resistance below $T_c$ for δ = 0.00 is smaller than that for δ = +0.12. For example, the resistance at $T$ = 43 K at 29.2 GPa for δ = 0.00 is 95% of that at $T_c$, whereas at 26.1 GPa and $T$ = 2 K, it is 83% of that at $T_c$ for δ = +0.12. Although the reduction in resistance does not always represent the superconducting volume fraction, one might infer that the fraction is smaller for δ = 0.00. However, it is important to note that the reduction strongly depends on the sample configuration in the DAC, which is hard to control precisely, although $T_c$ is less affected. In fact, the resistance below $T_c$ decreased significantly for δ = +0.01, although the pressure dependence of $T_c$ was almost the same as that for δ = 0.00 up to the maximum pressure measured of 35 GPa.

The $R$—$T$ curves at various pressures for δ = −0.16 and δ = −0.50 are shown in Fig. 5. The resistance for δ = −0.16 is several orders of magnitude higher than that for δ = 0.00, indicating that the insulating nature is intrinsic. Nevertheless, the resistance does not follow Arrhenius-type behavior but agrees well with variable-range hopping (VRH) behavior, at least above approximately 100 K ($T^{1/4}$ = 3.16), as it can be reproduced by the equation $R \propto \exp(T_0/T)^{1/4}$, where $T_0$ is a parameter representing the degree of carrier localization. This is consistent with the fact that the Ni valence is incommensurate with the crystal structure, suggesting that Anderson localization caused by partially removed oxygen ions is likely the origin of the insulating nature. The $T_0$ value decreases with increasing pressure, as shown in the inset of Fig. 5b, indicating that carrier localization becomes weaker at higher pressures.

The resistance for δ = −0.50 is significantly higher than that for δ = −0.16, showing Arrhenius-type behavior with the energy gap of Δ ~ 1900 K above approximately 200 K. This aligns with the expectation that $La_3Ni_2O_{6.5}$ is a Mott insulator with only divalent Ni ions, as predicted by electronic band structure calculations.[36] Insulating behavior has also been reported for δ = −0.65[24] and δ = −0.55.[37] For the former, VRH behavior was observed, while for the latter, Arrhenius-type behavior was assumed. The VRH behavior seems reasonable because the Ni valence in the compound is far from two. For δ = −0.55, one might question the validity of Arrhenius-type behavior at high pressures, as suggested by the curvature of the $R$—$T$ curves and the small gap values relative to the rather high measurement temperatures. Despite these concerns, the potential issue does not affect our conclusion because the chemical composition for $\delta = -0.55$ is different from $La_3Ni_2O_{6.5}$. In general, Mott insulating state is sensitive to the band filling.[38, 39]

The above-mentioned experimental findings result in the phase diagram shown in Fig. 6. Clearly, superconductivity appears adjacent to the insulating region, where Anderson localization occurs. The $T_c$ values show a stronger dependence on $\delta$ as the pressure increases. This observation can help narrow down the possible mechanisms of superconductivity and suggests that higher $T_c$ values may be achievable in certain areas of the phase diagram. For example, at a slightly smaller $\delta$ than 0.00, $T_c$ may be higher than 80 K at pressures significantly greater than 60 GPa. Although the Anderson localization tends to occur below $\delta = 0.00$, it weakens and the metallic nature is expected to appear under high pressures, as indicated by the pressure dependence of $T_0$ shown in Fig. 5*b*.

## 3. Conclusions

By preparing samples of $La_3Ni_2O_{7+\delta}$ ($\delta = -0.50, -0.16, 0.00, +0.01$, and $+0.12$), we established a comprehensive electronic phase diagram showing $T_c$ as a function of $P$ and $\delta$, by measuring resistance under high pressures. For $\delta = -0.50$, a Mott insulating state appears, which is destroyed by adding oxygen, leading to Anderson localization. For $\delta = 0$ or higher, superconductivity appears above approximately $P = 25$ GPa. Although $T_c$ remains almost constant against $\delta$ at 25 GPa or slightly higher, the pressure dependence of $T_c$ becomes stronger with increasing $\delta$. Consequently, $T_c$ rapidly decreases with increasing pressure for larger $\delta$ values.

**Acknowledgment**

We would like to express our gratitude to Professors K. Kuroki (Osaka Univ.), H. Sakakibara (Tottori Univ.), and M. Ochi (Osaka Univ.), and Dr. M. Kohno (NIMS) for fruitful discussion. This research was conducted in part using the AtomWorks-Adv.[40] provided by the Materials Data Platform of NIMS, and was supported by World Premier International Research Center Initiative (WPI), MEXT, Japan, and by JSPS KAKENHI Grants No. JP20H05644 and JP24K01333. Figure 1 was drawn using VESTA.[41]

*E-mail: UEKI.Yuta@nims.go.jp

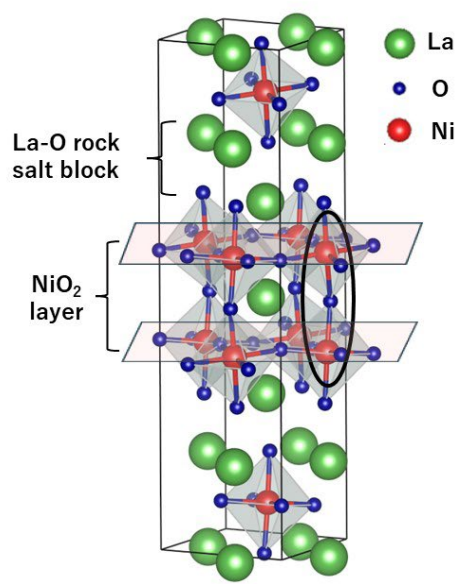

Fig. 1  Crystal structure of $La_3Ni_2O_7$. The red, blue, and green circles represent Ni, O, and La atoms, respectively. The box and circle indicate the unit cell and a pair of Ni atoms (see the text), respectively.

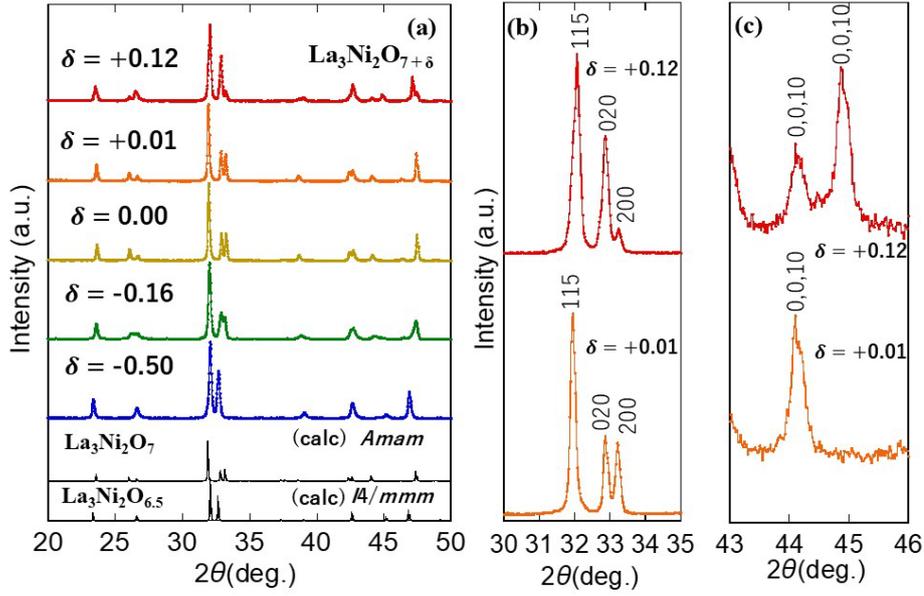

Fig. 2 Powder XRD patterns of the samples of La$_3$Ni$_2$O$_{7+\delta}$ ($\delta$ = −0.50, −0.16, 0.00, +0.01, and +0.12), and the calculated patterns for La$_3$Ni$_2$O$_7$ with the orthogonal and tetragonal symmetries of space groups *Amam* and *I4/mmm*, respectively (*a*). Panels *b* and *c* magnify the XRD patterns for $\delta$ = +0.12 and +0.01 between 30° and 35°, and between 43° and 46°, respectively.

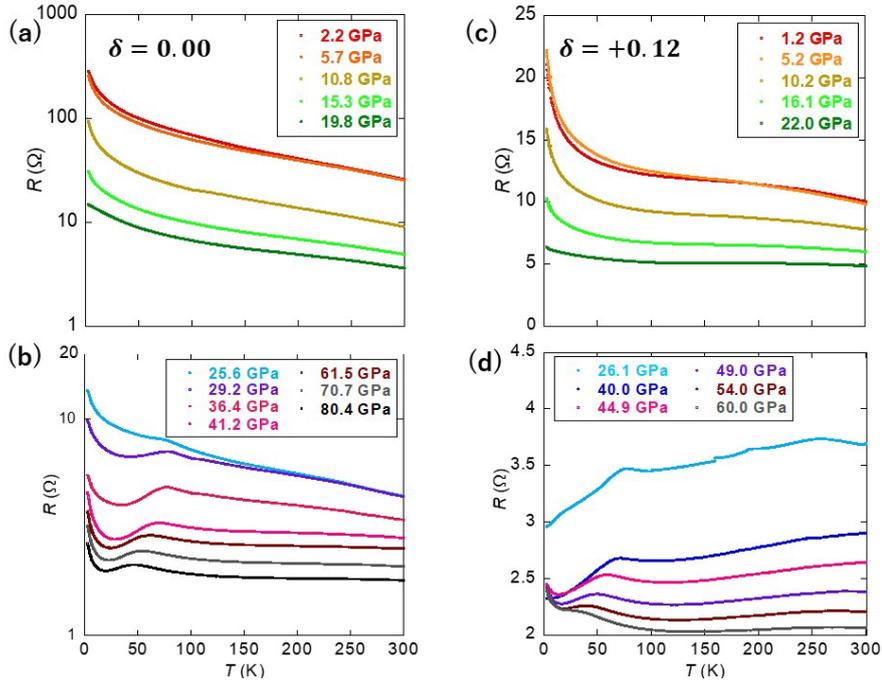

Fig. 3 Temperature dependence of electrical resistance for $\delta$ = 0.00 (*a* and *b*) and $\delta$ = +0.12 (*c* and *d*) under various pressures.

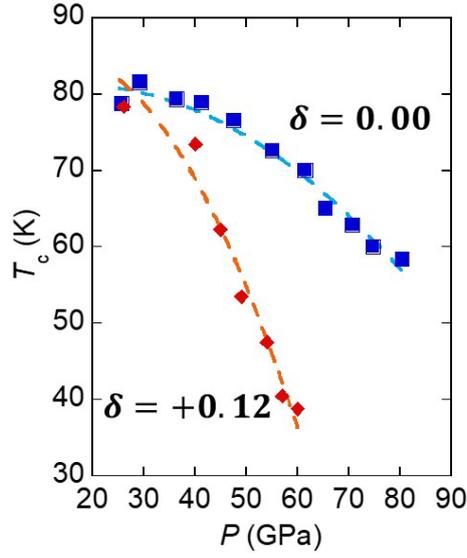

Fig. 4  Pressure dependence of $T_c$ estimated for $\delta = 0.00$ (red diamonds) and $\delta = +0.12$ (blue squares). The lines are guides for the eye.

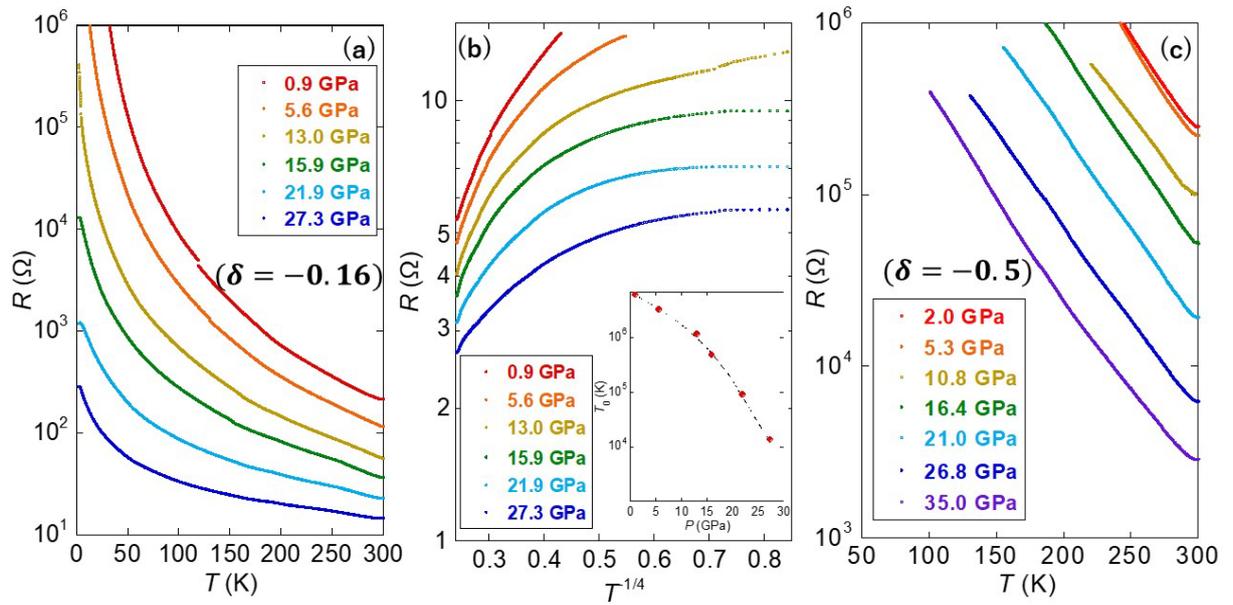

Fig. 5  Electrical resistance under various pressures for $\delta = -0.16$ as a function of $T$ (*a*) and $1/T^{1/4}$ (*b*), and for $\delta = -0.50$ (*c*). The inset in the panel *b* shows the pressure dependence of $T_0$ estimated from the data between approximately 100 K and 300 K using the equation $R \propto \exp(T_0/T)^{1/4}$.

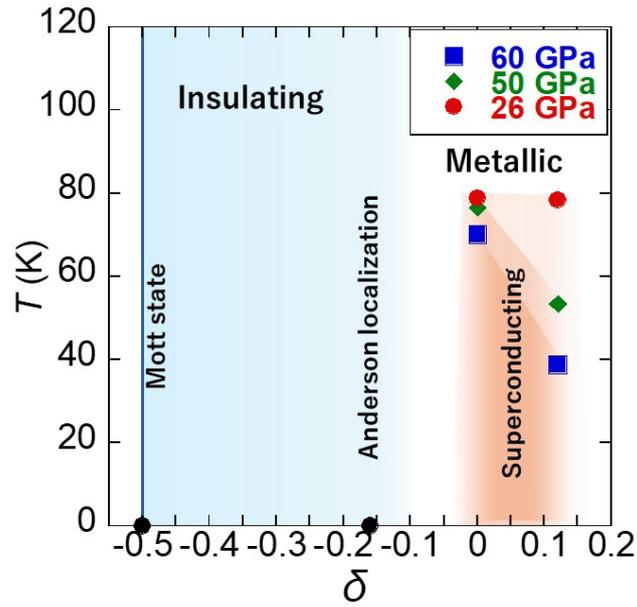

Fig. 6  The $T$—$\delta$ phase diagrams under approximately 26 GPa, 50 GPa, and 60 GPa. The three phase diagrams for the pressures are depicted overlapping each other. The $T_c$ under the pressures are shown by the red circles, green diamonds, and blue squares, respectively. The black squares on the $\delta$ axis show the $\delta$ position for the samples used. The colored areas are guides for the eye.